
\input psbox.tex
\splitfile{\jobname}
\let\autojoin=\relax
\magnification=1200

\input psbox.tex

\vsize=8.75truein
\baselineskip=12pt

\def\etal{{\it et al.}}
\def\tphi{\tilde\phi}
\def\pr{Phys.\ Rev.}
\def\prl{Phys.\ Rev.\ Lett.}
\def\la{\langle}
\def\ra{\rangle}
\def\half{\hbox{$\textstyle {1\over 2}$}}

\def\ve{\vec e}
\def\vI{\vec I}
\def\vj{\vec j}
\def\vk{\vec k}
\def\vq{\vec q}
\def\vr{\vec r}
\rightline{cond-mat/9305024}
\bigskip

\centerline{\bf Theories for $\nu=1/2$ in Single- and Double-Layer Systems}
\bigskip
\centerline{Bertrand I. Halperin}
\centerline{Physics Department}
\centerline{Harvard University}
\centerline{Cambridge, MA \ 02174}
\bigskip
\bigskip

\noindent
Manuscript for an invited talk at the Tenth International Conference on
Electronic Properties of Two-Dimensional System, Newport, RI, May~31--June~4,
1993. (Proceedings to appear in {\it Surface Science}.)
\bigskip

\noindent
\hskip 1truein\vbox{\hsize=4.5truein
\centerline{\bf Abstract}
\bigskip

\noindent
Recent experiments have shown that a quantized Hall plateau can occur
in double layer systems at total filling factor $\nu=1/2$, though there
is no plateau at $\nu=1/2$ in a normal single layer system. For the single
layer system, considerable insight has been provided by a theory based on the
fermion Chern-Simons picture, where the electrons are transformed into
fermions that carry two flux quanta of a Chern-Simons gauge field. A
similar picture can be used to characterize ground states which have
been proposed for the two layer system.

}
\bigskip

\noindent
{\bf I. Introduction}
\bigskip

During the course of the past few years, as experiments have continued to
reveal the structure of electronic states in a partially filled Landau
level, a variety of theoretical approaches have been developed to understand
these systems. One of the most useful of these approaches employs a
singular gauge transformation to convert the electrons to a system of
particles interacting with a Chern-Simons gauge field [1--11].
In this description, a flux tube
containing an integer number $\tphi$ of quanta of the Chern-Simons
magnetic field is attached to each particle. If $\tphi$ is an even
integer~[5--11], then the transformed particles obey Fermi statistics.
The motivation for
employing this singular gauge transformation is that for various rational
values of the Landau-level
filling factor $\nu$, with an appropriate choice of $\tphi$,
if one treats the transformed system in a simple Hartree approximation, the
resulting ground state is nondegenerate, and therefore has a reasonable
chance of being a good first approximation to the true ground state of the
system. Moreover, one may hope to calculate corrections to the ground state
and study the dynamic response of the system by using standard techniques
of diagrammatic perturbation theory, beginning with the Hartree
ground state~[9, 10, 12].

As an important example, if $\nu=p/(2p+1)$, where $p$ is a positive or
negative integer, and if we choose $\tphi=2$, then the mean-field
ground state of the transformed system is a collection of fermions in an
{\it effective} magnetic field whose strength is such that exactly
$|p|$ Landau levels are filled by fermions. The ground state is, therefore,
stabilized by an energy gap separating it from the excited states. This
provides a natural explanation for the most prominent fractional quantized
Hall states, which are observed
at these filling fractions. At the mean-field level, the
fermion Chern-Simons description is essentially equivalent to Jain's
composite fermion description of these quantized Hall states~[8].

In a recent paper (HLR), P.A. Lee, N. Read, and the present author employed
the fermion Chern-Simons method to analyze the properties of a single layer
system, at $\nu=1/2$ and at various other even fractions, where the quantized
Hall effect has not been observed~[10].
At $\nu=1/2$, if one chooses $\tphi=2$,
the average Chern-Simons field just cancels the external magnetic field,
so that the Hartree ground state is just a filled Fermi sea of particles,
in zero magnetic field, with Fermi wavevector
$k_F=(4\pi n_e)^{1/2}$. (Here $n_e$
is the areal density of the electrons. We assume that the electron spins
are fully aligned by the Zeeman field.)
Although there is no energy gap in this case, the density of states for
low energy particle-hole excitations is small, so that there is reason to
hope that the mean-field ground state may be stable with respect to the
particle-particle interactions, similar to the case of an ordinary Fermi
liquid. The detailed analysis of HLR gives rise to predictions for various
properties of the $\nu=1/2$ system, which seem to be in excellent qualitative
agreement with experiments and with exact calculations of finite systems.
The most striking of these predictions is an explanation for the surface
acoustic wave anomalies observed by Willett and coworkers~[13, 14]. A summary
of the most important results of the fermion Chern-Simons theory in the
single layer system will be given in Section~II below.

As is now well known, a quantized Hall plateau at total filling $\nu=1/2$ has
recently been observed in certain double layer systems by groups at
Princeton and Bell Laboratories~[15, 16].
A plateau at filling fraction $\nu=5/2$ was
observed earlier in single layer systems by Willett \etal~[17].
Although various explanations for these states have been advanced, there
remains a considerable amount of debate about the precise form of the ground
state in various cases.  I am not able to settle these questions, but
I will try to outline in Section~III below how the various postulated
quantized Hall states may be at least formulated in terms of the
fermion Chern-Simons picture as states with various forms of BCS pairing
among the particles near the Fermi surface. This suggests a simple phase
diagram for how the various states may be connected.

\bigskip

\noindent
{\bf II. The Single Layer System}
\bigskip

We summarize here some of the key results of the analysis of HLR~[10]
for a fully polarized single-layer system at $\nu=1/2$.

The density and current response functions have been obtained using the
Random Phase Approximation (RPA) or time-dependent Hartree approximation.
Here the transformed fermions are treated as free particles which respond
to the self-consistent Chern-Simons electric and magnetic fields
$\la\ve(\vr,t)\ra$ and $\la b(\vr,t)\ra$, as well as to the external
electromagnetic field and the self-consistent Coulomb potential of thee
particles~[9, 10, 12]. The equations for $\la b\ra$ and $\la\ve\ra$ are
$$
\eqalignno{
\la b\ra\,&=2\pi\tphi\la\rho\ra\;,&(2.1)\cr
\la\ve\ra\,&=-2\pi\tphi\hat z\times\la\vj\ra\;,&(2.2)\cr}
$$
where $\la\rho\ra$ and $\la\vj\ra$ are the particle density and current,
respectively, and $\tphi=2$. The system is found to be ``compressible'' at
long wavelengths, which means more precisely that the static density
response function $\chi_{\rho\rho}(q)$ is determined by the diverging
Coulomb interaction for $q\to0$:
$$
\chi_{\rho\rho}(q)^{-1}\sim v(q)=
{2\pi e^2\over\epsilon q}\;.
\eqno(2.3)
$$
The frequency-dependent density response function $\chi_{\rho\rho}(q,\omega)$
has, in addition to the pole at the cyclotron frequency that exhausts the
$f$-sum rule for $q\to0$, a diffusive pole at a low frequency $\omega=
- -i\gamma_q$ which we write in the form
$$
\gamma_q=q^2v(q)\sigma_{xx}(q)\;,
\eqno(2.4)
$$
where $\sigma_{xx}(q)$ is the wavevector-dependent longitudinal
conductivity (we assume $\vq\,\|\,\hat x$).
According to the RPA, for a system without impurities,
$\sigma_{xx}(q)$ is given by~[10]
$$
\sigma_{xx}(q)={e^2\over8\pi\hbar}\,{q\over k_F}\;.
\eqno(2.5)
$$

More generally, if impurity scattering is taken into account, we expect that
(2.5) applies for $q\gg\ell^{-1}$, where $\ell$ is the transport
mean-free path at $\nu=1/2$. For $q\to0$, the conductivity goes to a finite
value which may be obtained by replacing $q$ on the right hand side of
(2.5) by $(2\ell^{-1})$. The value of $\ell$ is expected to be much
smaller than the transport mean free path in zero magnetic field. This
is because the dominant mechanism for scattering of carriers at $\nu=1/2$
comes from static fluctuations of the Chern-Simons magnetic field due to
inhomogeneities in the electron density induced by random variations
in the density of charged impurities in the doping layer, a
mechanism which does not occur
for electrons in zero magnetic field~[18]. A crude estimate of $\ell$, at
$\nu=1/2$, was obtained by assuming that the charged impurities are
uncorrelated within the doping layer, and are equal in number to the
electrons in the conducting layer. If scattering is treated in the
Born approximation, one finds a value of $\ell$ which is just equal
to the setback distance $d_s$ of the doping layer in this model~[10].
Experiments suggest that our crude estimate for $\ell$ is about a factor of
three smaller than the actual values in the highest mobility samples~[14, 19].

An important effect arising from dynamic fluctuations of the Chern-Simons
vector potential is a large renormalization of the effective mass of the
transformed fermions. If the bare mass is small, so that the cyclotron
energy is large compared to the scale of the electron-electron interactions,
then the effective mass becomes independent of the bare mass, and is determined
by the electron-electron interaction. Using a self-consistent analysis
based on the leading diagrams in perturbation theory, HLR propose that
there is a logarithmic divergence of the effective mass at the Fermi
energy for Coulomb interactions, and a stronger power-law divergence
for short range interactions, but that the most essential features of Fermi
liquid theory are preserved in either case.
Note that expressions (2.3)--(2.5) for the density response function and the
conductivity are independent of the electron mass, and we believe that they
are not affected by the divergent mass renormalization. (The results for
the mean-free path in the presence of impurities are also independent of the
electron mass.)

One place where the effective mass enters directly is in the
expression of HLR for the energy gaps $E_g^{(\nu)}$ for the principal
quantized Hall states at $\nu=p/(2p+1)$. For an interaction that behaves like
$e^2/\epsilon r$ at large distances, HLR predict the following asymptotic
form for the energy gap at large $p$:
$$
E_g^{(\nu)}\sim{4\over\pi}\,{e^2\over\epsilon\ell_0}\,{1\over D(\ln D+C)}\;,
\eqno(2.6)
$$
where $D=|2p+1|$ is the denominator of the fraction and $C$ is a constant
which depends on the short distance behavior of the potential.
(This formula is based on a self-consistent analysis of the leading
correction to the quasiparticle self energy arising from interactions
with fluctuations in the transverse gauge field; it is possible that
it may be modified by other singular contributions.)
A good fit to numerical estimates~[20]
of the energy gaps at $\nu=1/3$, $2/5$, and $3/7$,
for a pure Coulomb interaction, may be obtained by choosing $C\approx2.5$
in that case.
The effects of finite layer thickness and inter-Landau-level mixing, which
occur in any real sample, would tend to increase the value of $C$ still
further. An energy gap of the form (2.6), with a relatively large value of
$C$, also gives a good fit to the data of Du \etal~[21], provided that one
accepts the proposal of those authors that the effects of impurity scattering
may be taken into account by subtracting a constant $\Gamma$, independent
of $\nu$, from the theoretical energy gap.

The linear wavevector dependence of $\sigma_{xx}(q)$, predicted by (2.5)
for $\nu=1/2$, is just what is needed to explain the anomalous surface
acoustic wave propagation, seen at short wavelengths by Willett \etal~[14]
The absolute values of $\sigma_{xx}(q)$ extracted by Willett \etal\ from
their data are larger than the theoretical values obtained from (2.5),
however, by a factor of $\approx2$. The theory of HLR also predicts that
the width of the anomaly should depend linearly on $q$ as the magnetic
field is varied away from the field corresponding to $\nu=1/2$.
This is in good agreement with the experimental observations.

Quasiparticle states for the transformed fermions which lie close to the
Fermi energy should not have a significant overlap with the wavefunction of a
single electron added to the ground state of a $\nu=1/2$ system.  A recent
analysis by He, Platzman and Halperin~[22], building on the results of HLR,
suggests that the spectral density $A(\omega)$ for the electron Green's
function vanishes as $e^{-\omega_0/|\omega|}$, for $|\omega|\to0$,
where $\omega_0$ is a constant.  Following this analysis, they predict a
pseudogap in $A(\omega)$, which is in reasonable agreement with recent
tunneling experiments~[23].

The general methods of HLR can be applied to various other even-denominator
fractions, including $\nu=1/4$, $3/4$, $3/2$, $3/8$, etc.
Chklovskii and Lee have shown, however, that a more sophisticated analysis
is necessary to understand the value of the electrical conductivity
$\sigma_{xx}$ at the higher order even fractions, because the Born
approximation for scattering becomes quite poor in this situation~[24].

\vfill\eject
\bigskip

\noindent
{\bf III. Double Layer Systems}
\bigskip

As a model to describe a double layer system, we shall introduce an ``isospin''
index $\tau=\pm 1$, which distinguishes between the two layers, in addition
to the position $\vr$ in the $x$-$y$ plane. The Coulomb interaction
between two electrons then has different forms $V_{++}(\vr-\vr^{\,\prime})$ and
$V_{+-}(\vr-\vr^{\,\prime})$, depending on whether the two electrons
are in the same or in different layers~[25]. In the simplest case where each
separate layer is considered to be of zero thickness, we may write
$$
\eqalignno{
V_{++}(r)\,&=e^2/\epsilon r\;,&(3.1)\cr
V_{+-}(r)\,&= e^2/\epsilon (r^2+d^2)^{1/2}\;,&(3.2)\cr}
$$
where $d$ is the separation between the layers. In addition, we
introduce a term to represent tunneling between the layers, which we
write as
$$
H_t=-tI_x\;,
\eqno(3.3)
$$
where $t$ is the tunneling matrix element, and $I_x$ is the $x$ component
of the total isospin operator $\vI$. We assume that the actual spins of the
electrons are completely polarized in the direction of the magnetic field,
and we consider only the case where there is a mirror symmetry between the
two layers. In our discussions we consider that the system employed in
[15], consisting of a single wide quantum well in which the self-consistent
Coulomb potential creates a barrier in the middle of the well, with
maxima in the electron density at the two edges, is equivalent to a double
layer system with a relatively large value of the tunneling matrix element
$t$.

We shall limit our discussions here to the case where the {\it total} filling
factor $\nu$ is equal to $1/2$; i.e.\ there is a total of one electron per
flux quantum in the two layers combined. Then, if the system is confined
to the lowest Landau level, there are essentially two dimensionless
parameters in our model $\tilde d\equiv d/\ell_0$, and $\tilde t\equiv t/
(e^2/\epsilon\ell_0)$.

Let us first consider the case where $\tilde d=0$, so that $V_{++}=V_{+-}$.
If $\tilde t$ is also equal to zero, then the Hamiltonian $H_0$ possesses
full $SU(2)$ symmetry in the isospin $\vI$. In fact, $H_0$ is
equivalent to the Hamiltonian for a single layer system with two spin
states and no Zeeman term to split the degeneracy.
The simplest assumption (though not universally believed~[26]) is
that the ground state of the single layer system would be completely
polarized at $\nu=1/2$, even in the absence of Zeeman interactions. If this
is the case, then for the two layer system with $\tilde d=0$, the
effect of $H_t$, for any positive value of $\tilde t$, is simply to align
the isospin polarization in the $x$-direction.
Specifically, this means that every electron is restricted to the isospin
state $I_x=1/2$, i.e., the lowest subband, which is
the even combination of states in the
two layers. Since the Hamiltonian is equivalent to that of a fully polarized
single layer system, we expect, as discussed in Section~II above, that the
ground state can be described by gauge transformed fermions with a single
Fermi surface, having $k_F=(4\pi n_e)^{1/2}=\ell_0^{-1}$, and no
quantized Hall effect.

Let us now consider the case where $\tilde d$ is nonzero and $\tilde t$ is
infinite. Every electron must have $I_x=1/2$, and hence all
electrons have the same interaction, $\bar V(r)=\half
[V_{++}(r)+V_{+-}(r)]$. If $\tilde d$ is very large, then $V_{+-}\approx0$,
and $\bar V (r)\approx\half\,V_{++}(r)$. Therefore, for large $\tilde d$,
the ground state is the same as for $\tilde d=0$,
and we expect to find a Fermi surface with no quantized Hall effect.
(The only change from $\tilde d=0$ is that the energy scale is reduced by
a factor of 2.)

According to the numerical calculations of Greiter, Wen and Wilczek~[27] for
a two layer system with $\tilde t=\infty$, there should exist an intermediate
range $\tilde d_{\min}<\tilde d<\tilde d_{\max}$, where a quantized Hall effect
does occur at $\nu=1/2$. Their calculations suggest that the quantized Hall
state has a very high overlap with the so-called Pfaffian state, originally
described by Moore and Read~[5], and further analyzed by
Greiter \etal~[7].  From the point
of view of the ground state symmetry, this state
can also be understood in terms of the fermion Chern-Simons picture as
a state where the fermions near to the Fermi surface are paired in a
BCS-like state, with orbital angular momentum $\ell_z=-1$, and isospin
$I_x=1$. (In terms of untransformed electrons, the state may be crudely
described as made up of pairs with angular momentum $\ell_z=1$, which
are then ``condensed'' into a Laughlin state of degree $m=8$.)
Moore and Read have suggested that the charged excitations of the
Pfaffian state have a different kind of statistics from what might be
expected in a simple pairing state, and perhaps there are other subtle
differences as well. We shall not distinguish here, however, between the
Pfaffian state and the Chern-Simons BCS state with pairing $\ell_z=-1$
and $I_x=1$.

Let us next consider the case $\tilde t=0$, $\tilde d\ne0$. Now, $I_z$
is a good quantum number of the system, and if there is equal population
of the two layers, the ground state must have $I_z=0$. (For $\tilde d\ne0$,
the Hamiltonian does not commute with $I_x$. The ground state has
$\la I_x\ra=0$, for $t=0$, but is not generally an eigenstate of $I_x$.)
In the limit $\tilde d\to\infty$, for $\tilde t=0$, the system becomes
two uncoupled layers, with $\nu=1/4$ in each layer. Experiments on single
layer systems show that there should be no quantized Hall effect in this
case~[14]. According to the theory of HLR, there should be a separate Fermi
surface of transformed fermions in each layer (seeing separate Chern-Simons
fields, with $\tphi=4$, in each layer), and a Fermi wavevector
$k_F=(2\pi n_e)^{1/2}=(2\ell_0^2)^{-1/2}$.

Numerical calculations for systems with $\tilde t=0$ again indicate that for
an intermediate range
$\tilde d_{\min}^{\,\prime}<\tilde d<\tilde d_{\max}^{\,\prime}$,
there should exist a quantized Hall plateau at $\nu=1/2$~[28--30].
The ground state
in this case has been found to have a high degree of overlap with the
so-called 331 state, first proposed in 1983 as a possible generalization of
Laughlin's wavefunctions to an even denominator fraction~[31]. The 331 state
has been characterized by various authors as a system of two types of
fermion with a $2\times2$ matrix of Chern-Simons interactions~[32].
However, the state may also be characterized in the spirit of Greiter
\etal~[7] as a system of fermions coupled to a single Chern-Simons field
(with coupling strength $\tphi=2$), whose ground state has BCS pairing
with $\ell_z=-1$ and $I_z=0$.

What happens for intermediate values of $\tilde t$, when $\tilde d\ne0$?
A simple schematic phase diagram, compatible with our previous discussion,
is presented in Figure~1.

\topinsert
{\baselineskip=10pt
\centerline{\psbox{fig1.eps}}
\medskip

\noindent
{\bf Figure~1}. Schematic phase diagram for the ground state of a two layer
system at total filling $\nu=1/2$. Variables $\tilde d$ and $\tilde t$ are
respectively
the separation between layers, in units of the magnetic length $\ell_0$,
and the tunneling strength between layers, in units of $e^2/\epsilon\ell_0$.
Phase~$A$ has two essentially independent layers of filling factor
$\nu=1/4$, with a separate Fermi surface in each layer, and no quantized Hall
effect. Phases~$B$ and $B'$ behave like a single layer at $\nu=1/2$, with
electrons in the subband
which is an even combination of states in the two
layers. These phases have a single Fermi surface for the gauge transformed
fermions, and no quantized Hall effect. Phase~$C$ is a quantized Hall state
which evolves continuously as a function of $\tilde t$ from
 a state with the symmetry
of the ``331 state'' at $\tilde t=0$, to a state with the symmetry of the
``Pfaffian'' state at $\tilde t=\infty$.

}
\bigskip
\endinsert

The phase labeled $A$ consists of two essentially independent layers with
$\nu=1/4$ and a separate Fermi surface for transformed fermions in each layer.
Phases $B$ and $B'$ have a large value of $\la I_x\ra$, and contain a single
Fermi surface for transformed fermions with isospin $I_x=1/2$, the even
combination of states in the two layers.

The phase labeled $C$, which occurs for intermediate values of the parameter
$\tilde d$, is a quantized Hall state. Within the fermion Chern-Simons
picture, we characterize the entire phase as a state with a BCS gap at
the Fermi surface due to pairing in a state of isospin 1 and $\ell_z=-1$.
Specifically, we expect pairing of the form
$$
c_{\vk\tau}c_{\vk^{\,\prime}\tau'}\approx
Q(\vr)(k_x-ik_y)f_{\tau\tau'}e^{i(\vk+\vk^{\,\prime})\cdot\vr}\;,
\eqno(3.4)
$$
where $c_{\vk\tau}$ is the annihilation operator for a transformed fermion
with wavevector $\vk$ and isospin $\tau$,
the wavevectors $\vk$ and $\vk^{\,\prime}$ are
close to the Fermi surface at diametrically opposite points, and $Q(\vr)$
is an order paramter whose pair correlation function $\la Q^+(\vr)Q(\vr^{\,
\prime})\ra$
falls off at large separations as a power of $|\vr-\vr^{\,\prime}|$
(i.e., the system has ``quasi-long-range order'' in the ground state).
The matrix $f_{\tau\tau'}$ is symmetric in the isospin indices, and we
hypothesize that it varies continuously as a function of the tunneling
strength $\tilde t$ between the two limits:
$$
\eqalignno{
f_{\tau\tau'}\to \delta_{\tau,-\tau'}
\;,&\qquad{\rm for}\qquad \tilde t\to0\;,&(3.5)\cr
f_{\tau\tau'}\to 1
\;,&\qquad{\rm for}\qquad \tilde t\to\infty\;,&(3.6)\cr}
$$
corresponding to pairs with $I_z=0$ and $I_x=1$, respectively. We also expect
that the expectation value $\la I_x\ra$ for the {total} isospin of the
electron system should increase continuously from $\la I_x\ra=0$ at
$\tilde t=0$ to $\la I_x\ra=N/2$ (full polarization) at $\tilde t=\infty$.

Numerical calculations by He \etal~[29, 30] suggest that for the values of
$\tilde t$ and $\tilde d$ which correspond to the Princeton and Bell
Laboratories experiments, there is a high degree of overlap between
the ground state at $\nu=1/2$ and the 331 state, which has $I_z=0$.
Thus, it appears that there is only a small amount of $\la I_x\ra$
polarization, even for the Princeton experiment where $\tilde t$ is
relatively large.

The calculations of He \etal~[30] support the conjecture that a quantized
Hall state should exist for an intermediate range of separations $\tilde d$,
for any value of the parameter $\tilde t$.  The conjecture that the 331
state can be continuously connected with the quantized Hall state at
$\tilde t=\infty$ is also compatible with the observation by Greiter \etal~[7]
that the Pfaffian state is realized by taking the fully antisymmetric
part of the spatial portion of the 331 wavefunction.

We do not address here the nature of the phase transitions between
the various regions of Figure~1. Of course, we cannot exclude at this
stage the possibility that the actual phase diagram is more complicated, with
various other intermediate phases occurring. If our starting assumption,
that the ground state for $\tilde t=\tilde d=0$ has complete spontaneous
alignment of the isospin vector $\vI$, is {\it not} correct, then there must
be a more complicated phase structure than we have indicated near the lower
left corner of Figure~1. Among the theoretical possibilities for the ground
state at $\tilde t=\tilde d=0$ are the following: (1)~there might be an
isospin singlet ground state with some type of energy gap, which would thus
exhibit a quantized Hall effect; (2)~there might be an isospin singlet
ground state with no energy gap, described within the fermion Chern-Simons
picture as having a single Chern-Simons field, with $\tphi=2$, and
a Fermi surface, with $k_F=(2\pi n_e)^{1/2}$, for each isospin
state; or (3)~the 331 state might exist as a stable ground state all the
way down to the point $\tilde t=\tilde d=0$. Since the 331 state is not
an eigenstate of $I^2$, it cannot be the true ground state for a finite
system at $\tilde t=\tilde d=0$; however, it could be the ground state
of an infinite system if there is a spontaneously broken isospin symmetry.

We note that BCS pairing, in the fermion Chern-Simons picture with
$\tphi=2$, has also been used to discuss the spin-singlet ``hollow-core''
ground state of Haldane and Rezayi~[33],
originally proposed as an explanation
for the quantized Hall state of a single layer at $\nu=5/2$. (This is a state
where the lowest Landau level is completely full and there is a one-half
electron per flux quantum in the second Landau level.) In this case the
BCS pairing has $\ell_z=-2$ for the transformed fermions, corresponding
to pairs with $\ell_z=0$ for the original electrons~[5, 7].

\bigskip

\noindent
{\bf IV. Conclusion}
\bigskip

Although many details remain to be understood, it seems clear that the
transformation to fermions with a Chern-Simons field is a powerful tool
for understanding the behavior of electrons at $\nu=1/2$, in both single
and double layer systems.
\vfill\eject
\bigskip

\noindent
{\bf Acknowledgments}
\bigskip

This review has benefited greatly from discussions with P.A. Lee, N. Read,
S. He, R.H. Morf, F. Wilczek, P.M. Platzman, J.P. Eisenstein, and R.L. Willett.
The work has been supported in part by NSF grant DMR-91-15491.
\bigskip

\noindent
{\bf References}
\bigskip

\item{[1]} S.M. Girvin and A.H. MacDonald, \prl\ 58 (1987) 1252.

\item{[2]} S.-C. Zhang, H. Hanson, and S. Kivelson, \prl\ 62 (1989) 82;
\prl\ 62 (1989) 980(E); S.-C. Zhang, Int.\ J. Mod.\ Phys.\ B 6 (1992) 25;
D.-H. Lee and M.P.A. Fisher, \prl\ 63 (1989) 903.

\item{[3]} D.-H. Lee, S. Kivelson, and S.-C. Zhang, \prl\ 68 (1992) 2386;
S. Kivelson, D.-H. Lee, and S.-C. Zhang, \pr\ B 46 (1992) 2223.

\item{[4]} D. Schmeltzer, \pr\ B 46 (1992) 1591.

\item{[5]} G. Moore and N. Read, Nucl.\ Phys.\ B 360 (1991) 362.

\item{[6]} M. Greiter and F. Wilczek, Mod.\ Phys.\ Lett.\ B 4 (1990) 1063.

\item{[7]} M. Greiter, X.-G. Wen, and F. Wilczek,
\prl\ 66 (1991) 3205; Nucl.\ Phys.\ B 374 (1992) 567;
M. Greiter and F. Wilczek, Nucl.\ Phys.\ B 370 (1992) 577.

\item{[8]} J.K. Jain, \prl\ 63 (1989) 199; \pr\ B 40 (1989) 8079;
\pr\ B 41 (1990) 7653; J.K. Jain, S.A. Kivelson, and N. Trivedi,
\prl\ 64 (1990) 1297.

\item{[9]} A. Lopez and E. Fradkin, \pr\ B 44 (1991) 5246.

\item{[10]} B.I. Halperin, P.A. Lee, and N. Read, \pr\ B 47 (1993) 7312.

\item{[11]} V. Kalmeyer and S.-C. Zhang, \pr\ B 46 (1992) 9889.

\item{[12]} R.B. Laughlin, \prl\ 60 (1988) 2677;
A.L. Fetter, C.B. Hanna, and R.B. Laughlin, \pr\ B 39 (1989) 9679;
Q. Dai, J.L. Levy, A.L. Fetter, C.B. Hanna, and R.B. Laughlin,
\pr\ B 46 (1992) 5642.

\item{[13]} R.L. Willett, M.A. Paalanen, R.R. Ruel, K.W. West, L.N. Pfeiffer,
and D.J. Bishop, \prl\ 54 (1990) 112.

\item{[14]} R.L. Willett, R.R. Ruel, M.A. Paalanen, K.W. West, and
L.N. Pfeiffer, \pr\ B 47 (1993) 7344.

\item{[15]} Y.W. Suen, L.W. Engel, M.B. Santos, M. Shayegan, and D.C. Tsui,
\prl\ 68 (1992) 1379.

\item{[16]} J.P. Eisenstein, G.S. Boebinger, L.N. Pfeiffer, K.W. West,
and S. He, \prl\ 68 (1992) 1383.

\item{[17]} R.L. Willett, J.P. Eisenstein, H.L. Stormer, D.C. Tsui,
A.C. Gossard, and J.H. English, \prl\ 59 (1987) 1776.

\item{[18]} This scattering mechanism was also discussed by V. Kalmeyer and
S.-C. Zhang in Ref.~11.

\item{[19]} H.L. Stormer, K.W. Baldwin, L.N. Pfeiffer, and K.W. West,
Solid State Commun.\ 84 (1992) 95.

\item{[20]} N. d'Ambrumenil and R.H. Morf, \pr\ B 40 (1989) 6108.

\item{[21]} R.R. Du, H.L. Stormer, D.C. Tsui, L.N. Pfeiffer, and K.W. West,
preprint.

\item{[22]} S. He, P.M. Platzman, and B.I. Halperin, preprint.

\item{[23]} J.P. Eisenstein, L.N. Pfeiffer, and K.W. West,
\prl\ 69 (1992) 3804.

\item{[24]} D.B. Chklovskii and P.A. Lee, preprint.

\item{[25]} T. Chakraborty and P. Pietil\"ainen, \prl\ 59 (1987) 2784.

\item{[26]} L. Belkhir and J.K. Jain, preprint.

\item{[27]} M. Greiter, X.-G. Wen, and F. Wilczek, \pr\ B 46 (1992) 9586.

\item{[28]} D. Yoshioka, A.H. MacDonald, and S.M. Girvin,
\pr\ B 39 (1989) 1932.

\item{[29]} S. He, X.C. Xie, S. Das Sarma, and F.C. Zhang,
\pr\ B 43 (1991) 9339.

\item{[30]} S. He, S. Das Sarma, and X.C. Xie, \pr\ B 47 (1993) 4394.

\item{[31]} B.I. Halperin, Helv.\ Phys.\ Acta 56 (1983) 75.

\item{[32]} See, for example, D. Schmeltzer, preprint.

\item{[33]} F.D.M. Haldane and E. Rezayi, \prl\ 60 (1988) 956;
\prl\ 60 (1988) 1886(E).

\autojoin
\bye